\documentclass[12pt]{article}%
\usepackage{amssymb}
\usepackage{amsfonts}
\usepackage{amsmath}
\usepackage{graphicx}%
\begin{document}
\begin{titlepage}
\vspace{1.5cm}
\begin{center}
\
\\
{\bf\large A quark model study of  strong decays of  $X\left(
3915\right)  $ }
\\
\date{ }
\vskip 0.70cm
P. Gonz\'{a}lez
\vskip 0.30cm
{ \it Departamento de F\'{\i}sica Te\'orica -IFIC\\
Universitat de València-CSIC \\
E-46100 Burjassot (Valencia), Spain.} \\ ({\small E-mail:
pedro.gonzalez@uv.es})
\end{center}
\vskip 1cm \centerline{\bf Abstract}
Strong decays of $X\left(
3915\right)  $ are analyzed from two quark model descriptions of  $X\left(
3915\right)  $, a conventional one in terms of the Cornell
potential and an unconventional one from a Generalized Screened potential. We conclude that the experimental
suppression of the  OZI allowed decay $X\left(  3915\right)
\rightarrow D\overline{D}$  might be explained in both cases as due to the momentum dependence of
the decay amplitude. However, the experimental significance of the OZI forbidden decay $X\left(
3915\right)  \rightarrow\omega J/\psi$ could  favor an unconventional description.
\end{titlepage}

\section{Introduction \label{SI}}

In the last edition of the Review of Particle Physics \cite{PDG14} the former
$X\left(  3915\right)  $ charmonium state has been assigned to a conventional
$\chi_{c0}\left(  3915\right)  $ with quoted mass $M_{X(3915)}=3918.4\pm1.9$
$MeV$ and total width $\Gamma_{X(3915)}=20\pm5$ $MeV$. This assignment has
been a matter of controversy: in references \cite{Guo12,Ols15} it has been
argued that the mass, width, decay properties and production rates are
incompatible with a $\chi_{c0}\left(  2p\right)  $ state expected from
conventional descriptions as the ones provided by the Cornell model
\cite{Eic80} or the Godfrey-Isgur model \cite{GI85}. Indeed, alternative
descriptions, based on four quark structures -meson-antimeson molecule,
tetraquark, mixed charmonium-molecule...-, have been developed in the past
(for an extensive review of these alternatives see the recent report
\cite{Ch16} and references therein). In particular, some of the different
molecular like treatments \cite{Liu09,Bra09,Mo09,Li15} have allowed for the
calculation of masses as well as strong and electromagnetic decay properties
that can be compared to current data.

In this article we show that an unconventional description of the $X\left(
3915\right)  ,$ yet based on a quark-antiquark structure, as the one provided
by the Generalized Screened Potential Model (GSPM) \cite{Gon14, Gon15}, may
give better account of its decay properties than the conventional one from the
Cornell model. As a matter of fact the GSPM results, being closer to the ones
obtained from molecular like treatments, may provide a reasonable description
of data.

We shall centre first on the lack of evidence of the OZI allowed decay
$X\left(  3915\right)  \rightarrow D\overline{D}$. By using two different
decay models, $^{3}P_{0}$ and $C^{3\text{ }},$ for the calculation of the
amplitude, we shall show that the observed experimental suppression may be
explained either from the momentum dependence of the $^{3}P_{0}$ amplitude in
the case of the GSPM description of $X\left(  3915\right)  $ or from the
momentum dependence of the $C^{3\text{ }}$ amplitude in the case of the
Cornell description of $X\left(  3915\right)  $. Therefore no definite
conclusion about the conventional or unconventional nature of $X\left(
3915\right)  $ should be extracted from this decay. On the contrary, we shall
show that the significant partial width for the OZI suppressed decay $X\left(
3915\right)  \rightarrow\omega J/\psi$, which we shall analyze later, might
discriminate between both descriptions in favor of the GSPM one.

These contents are organized as follows. In Section \ref{SII} a comparative
description of $X\left(  3915\right)  $ with the Cornell potential versus the
Generalized Screened potential is presented. Then, in Section \ref{SIII} a
study of $X\left(  3915\right)  \rightarrow D\overline{D}$ with the $^{3}%
P_{0}$ and $C^{3}$ decay models is carried out for both descriptions of
$X\left(  3915\right)  $, the results being compared to the ones obtained from
other approaches. Section \ref{SIV} is dedicated to the analysis of the decay
$X\left(  3915\right)  \rightarrow\omega J/\psi$. Finally, in Section \ref{SV}
our main results and conclusions are summarized.

\section{$X\left(  3915\right)  $ quark model descriptions \label{SII}}

The Cornell potential \cite{Eic80}%
\begin{equation}
V_{Cor}(r)\equiv\sigma r-\frac{\zeta}{r}\text{ \ \ \ }\left(  r:0\rightarrow
\infty\right)  \label{Corpot}%
\end{equation}
with the parameters $\sigma$ and $\zeta$ standing for the string tension and
the color coulomb strength respectively, and refined models from it
\cite{GI85}, have been very successful in the description of the heavy
quarkonia spectra ($r$ standing for the quark-antiquark distance) below the
open-flavor two meson thresholds. Above these thresholds the effect of
two-meson channels have been explicitly implemented \cite{Eic04,Eic06} but a
good overall description of data seems difficult to be attained.

In Table~\ref{tab1}, from \cite{Gon15}, the calculated masses for $J^{++}$
Cornell charmonium sates (fifth column) from $V_{Cor}(r)$, with standard
effective parameters $\sigma=850$ MeV/fm, $\zeta=100$ MeV.fm , are listed
(these values provide a reasonable overall spectral description of charmonium
as well as bottomonium \cite{Gon14}). The value chosen for the charm mass
$m_{c}=1348.6$ MeV will be justified below.

\begin{table}[ptb]%
\begin{tabular}
[c]{cccccc}%
$J^{PC}$ & $%
\begin{array}
[c]{c}%
\text{GSPM}\\
\text{States}%
\end{array}
$ & $%
\begin{array}
[c]{c}%
M_{GSPM}\\
\text{MeV}%
\end{array}
$ & $%
\begin{array}
[c]{c}%
M_{PDG}\\
\text{MeV}%
\end{array}
$ & $%
\begin{array}
[c]{c}%
M_{Cor}\\
\text{MeV}%
\end{array}
$ & $%
\begin{array}
[c]{c}%
\text{Cornell}\\
\text{States}%
\end{array}
$\\\hline
&  &  &  &  & \\
$0^{++}$ & $1p_{\left[  T_{0},T_{1}\right]  }$ & $3456.1$ & $3414.75\pm0.31$ &
$3456.2$ & $1p$\\
$1^{++}$ & $1p_{\left[  T_{0},T_{1}\right]  }$ & $3456.1$ & $3510.66\pm0.07$ &
$3456.2$ & $1p$\\
$2^{++}$ & $1p_{\left[  T_{0},T_{1}\right]  }$ & $3456.1$ & $3556.20\pm0.09$ &
$3456.2$ & $1p$\\
&  &  &  &  & \\
$1^{++}$ & $2p_{\left[  T_{0},T_{1}\right]  }$ & $3871.7$ & $3871.69\pm0.17$ &
$3910.9$ & $2p$\\
&  &  &  &  & \\
$0^{++}$ & $1p_{\left[  T_{1},T_{2}\right]  }$ & $3897.9$ & $3918.4\pm1.9$ &
$3910.9$ & $2p$\\
&  &  &  &  & \\
$2^{++}$ & $2p_{\left[  T_{0},T_{1}\right]  }$ & $3903.0$ & $3927.2\pm2.6$ &
$3910.9$ & $2p$\\
&  &  &  &  & \\
$1^{++}$ & $1p_{\left[  T_{1},T_{2}\right]  }$ & $4017.3$ &  &  & \\
&  &  &  &  & \\
$0^{++}$ & $1p_{\left[  T_{3},T_{4}\right]  }$ & $4140.2$ &  &  & \\
&  &  &  &  & \\
&  &  & $X\left(  4140\right)  $ &  & \\
&  &  &  &  & \\
$2^{++}$ & $1p_{\left[  T_{1},T_{2}\right]  }$ & $4140.2$ &  &  & \\
&  &  &  &  & \\
&  &  &  &  & \\
&  &  &  &  & \\
$0^{++}$ & $1p_{\left[  T_{4},T_{5}\right]  }$ & $4325.1$ & $X\left(
4350\right)  $ & $4294.6$ & $3p$\\
&  &  &  &  &
\end{tabular}
\caption{Calculated $J^{++}$ charmonium masses, up to $4350$ MeV, from the
Cornell potential $V_{Cor}:$ $M_{Cor},$ and from the Generalized Screened
potential $V(r):M_{GSPM}$, with $\sigma=850$ MeV/fm, $\zeta=100$ MeV.fm and
$m_{c}=1348.6$ MeV. For the GSPM the $0^{++}\left(  1p_{\left[  T_{2}%
,T_{3}\right]  }\right)  $ row has been omitted since there is no bound state
in that energy region; for $1^{++}$ we do not list any state above $4080$ MeV
due to the current incomplete knowledge about thresholds above this energy;
the same for $2^{++}$ states above $4224$ MeV. Masses for experimental
resonances, $M_{PDG},$ have been taken from \cite{PDG14} (when a resonance
appears in the Particle Listing section of \cite{PDG14} but not in the Summary
Table we write the name of the resonance that contains the nominal mass
between parenthesis). For $p$ waves we quote separately the $np_{0}$, $np_{1}$
and $np_{2}$ states. }%
\label{tab1}%
\end{table}We shall focus our attention on $X\left(  3915\right)  $, a
$0^{+}(0^{++})$ charmonium state above the first corresponding $0(0^{++})$
threshold, $D\overline{D}$, at $3730$ MeV. In the Cornell model it should be
assigned to the $2p$ state: $\chi_{c0}\left(  2p\right)  $. Although the
calculated $\chi_{c0,1,2}\left(  2p\right)  $ mass in Table~\ref{tab1}
($3910.9$ MeV) is close to the experimental one ($3918.4$ MeV) a more
stringent test of the model involving other observables should be done before
making any definite assignment. For this purpose we shall consider strong
decays for which there are some experimental information.

For the sake of comparison, an unconventional description from the so called
Generalized Screened Potential Model (GSPM) will be used. The GSPM
\cite{Gon14, Gon15} is based on an effective quark-antiquark static potential
$V\left(  r\right)  $ that implicitly incorporates threshold effects, in
particular color screening from meson-meson configurations. The model has been
applied to heavy quarkonia showing that a reasonable overall description of
$J^{++}$ resonances below and above thresholds and of $1^{--}$ resonances
quite below threshold is feasible (the choice of the mass $m_{c}=1348.6$ MeV
allows for a precise spectral description of $X(3872)$ as a $0\left(
1^{++}\right)  $ state). More precisely, $V\left(  r\right)  $ is obtained
through a Born-Oppenheimer approximation from the lattice results for the
energy of two static color sources (heavy quark and heavy antiquark) in terms
of their distance, $E\left(  r\right)  $, when the mixing of the quenched
quark-antiquark configuration with open flavor meson-meson ones is taken into
account \cite{Bal05}. By calling $M_{T_{i}}$ with $i\geq1$ the masses of the
physical meson-meson thresholds, $T_{i}$, with a given set of quantum numbers
$I(J^{PC})$, and defining $M_{T_{0}}\equiv0$ for a unified notation (note that
$T_{0}$ does not correspond to any physical meson-meson threshold), the form
of $V\left(  r\right)  $ in the different energy regions (specified as energy
interval subindices) reads:%

\begin{equation}
V_{\left[  M_{T_{0}},M_{T_{1}}\right]  }(r)=\left\{
\begin{array}
[c]{c}%
\sigma r-\frac{\zeta}{r}\text{ \ \ \ \ \ \ \ \ \ \ \ \ \ \ \ \ \ \ \ \ \ \ }%
r\leq r_{T_{1}}\\
\\
M_{T_{1}}-m_{Q}-m_{\overline{Q}}\text{ \ \ \ \ \ \ \ \ \ }r\geq r_{T_{1}}%
\end{array}
\right.  \label{GSP1}%
\end{equation}
and%
\begin{equation}
V_{\left[  M_{T_{j-1}},M_{T_{j}}\right]  }(r)=\left\{
\begin{array}
[c]{c}%
M_{T_{j-1}}-m_{Q}-m_{\overline{Q}}\text{ \ \ \ \ \ }r\leq r_{T_{j-1}}\\
\\
\sigma r-\frac{\zeta}{r}\text{\ \ \ \ \ \ \ \ \ \ \ }r_{T_{j-1}}\leq r\leq
r_{T_{j}}\\
\\
M_{T_{j}}-m_{Q}-m_{\overline{Q}}\text{ \ \ \ \ \ \ \ \ \ }r\geq r_{T_{j}}%
\end{array}
\right.  \label{GSP2}%
\end{equation}
for $j>1,$ where $m_{Q}$ $\left(  m_{\overline{Q}}\right)  $ stands for the
mass of the heavy quark\ (antiquark) and with the crossing radii $r_{T_{i}}$
$\left(  i\geq1\right)  $ defined by the continuity of the potential as
\begin{equation}
\sigma r_{T_{i}}-\frac{\zeta}{r_{T_{i}}}=M_{T_{i}}-m_{Q}-m_{\overline{Q}}
\label{GSP3}%
\end{equation}

\begin{figure}
[ptb]
\begin{center}
\includegraphics[
height=2.3808in,
width=3.659in
]%
{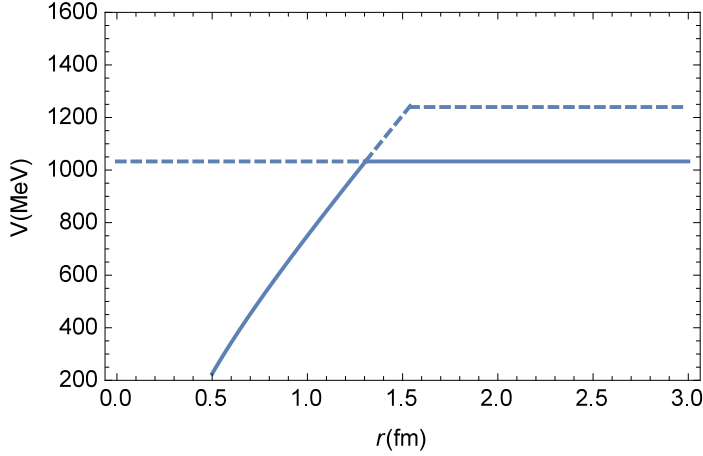}%
\caption{Generalized screened potential $V(r).$ The solid (dashed) line
indicates the potential in the first (second) energy region for $0^{+}%
(0^{++})$ $c\overline{c}$ states with $m_{c}=1348.6$ MeV, $\sigma=850$ MeV/fm,
$\zeta=100$ MeV.fm, $M_{T_{1}}=3730$ MeV ($r_{T_{1}}=1.31$ fm) and $M_{T_{2}%
}=3937$ MeV ($r_{T_{2}}=1.54$ fm).}%
\label{charm850pot0}%
\end{center}
\end{figure}

Thus, $V\left(  r\right)  $ has in each energy region between neighbor
thresholds a Cornell form but modulated at short and long distances by these
thresholds. Thus, for example in Fig. \ref{charm850pot0} the form of $V\left(
r\right)  $ in the\ first and second energy regions is drawn for
$c\overline{c}$ states with $I^{G}(J^{PC})=0^{+}(0^{++})$ quantum numbers,
whose first threshold $T_{1}$ corresponds to $D^{0}\overline{D}^{0}$ and its
second threshold $T_{2}$ to $D_{s}^{+}D_{s}^{-}$.%
%TCIMACRO{\FRAME{ftbpFU}{3.659in}{2.3808in}{0pt}{\Qcb{Generalized screened
%potential $V(r).$ The solid (dashed) line indicates the potential in the first
%(second) energy region for $0^{+}(0^{++})$ $c\overline{c}$ states with
%$m_{c}=1348.6$ MeV, $\sigma=850$ MeV/fm, $\zeta=100$ MeV.fm, $M_{T_{1}}=3730$
%MeV ($r_{T_{1}}=1.31$ fm) and $M_{T_{2}}=3937$ MeV ($r_{T_{2}}=1.54$ fm).}%
%}{\Qlb{charm850pot0}}{charm850pot0.eps}{\special{ language "Scientific Word";
%type "GRAPHIC";  maintain-aspect-ratio TRUE;  display "USEDEF";
%valid_file "F";  width 3.659in;  height 2.3808in;  depth 0pt;
%original-width 3.6115in;  original-height 2.341in;  cropleft "0";
%filename 'charm850pot0.eps';file-properties "XNPEU";}} }%
%BeginExpansion
%EndExpansion

\begin{figure}
[ptb]
\begin{center}
\includegraphics[
height=2.4396in,
width=3.659in
]%
{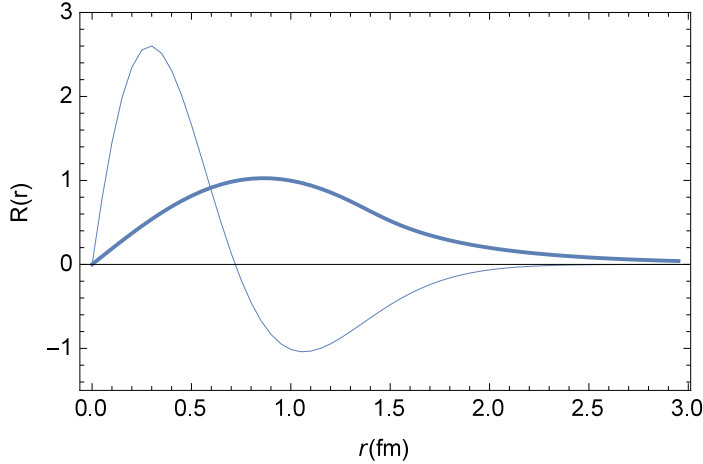}%
\caption{Radial wave functions R(r) (in units $fm^{-\frac{3}{2}}$) for the
$0^{++}\left(  1p_{\left[  T_{1},T_{2}\right]  }\right)  $ GSPM state (thick
line) and the $0^{++}\left(  2p\right)  $ Cornell state (thin line).}%
\label{wfx3915}%
\end{center}
\end{figure}

From (\ref{GSP1}) it is clear that the description of states far below the
lowest threshold $M_{T_{1}}$ is going to be identical to the Cornell one;
however, a completely different description of the states above $M_{T_{1}}$
comes out. For instance, the $0^{+}(0^{++})$ bound state in the energy region
$\left[  M_{T_{1}=D^{0}\overline{D}^{0}},M_{T_{2}=D_{s}^{+}D_{s}^{-}}\right]
$ is obtained by solving the Schr\"{o}dinger equation for $V_{\left[
M_{D^{0}\overline{D}^{0}},M_{D_{s}^{+}D_{s}^{-}}\right]  }(r)$. This
$1p_{\left[  T_{1},T_{2}\right]  }$ GSPM state with mass $3897.9$ MeV which
should be assigned to $X(3915),$ see Table~\ref{tab1}, differs greatly from
the $2p$ Cornell one as can be checked in Fig. \ref{wfx3915}, from
\cite{Gon15}, where the respective radial wave functions are plotted.%
%TCIMACRO{\FRAME{ftbpFU}{3.659in}{2.4396in}{0pt}{\Qcb{Radial wave functions
%R(r) (in units $fm^{-\frac{3}{2}}$) for the $0^{++}\left(  1p_{\left[
%T_{1},T_{2}\right]  }\right)  $ GSPM state (thick line) and the $0^{++}\left(
%2p\right)  $ Cornell state (thin line).}}{\Qlb{wfx3915}}{wfx3915.eps}%
%width 3.659in;  height 2.4396in;  depth 0pt;  original-width 3.6115in;
%original-height 2.3981in;  cropleft "0";  croptop "1";  cropright "1";
%cropbottom "0";  filename 'wfx3915.eps';file-properties "XNPEU";}} }%
%BeginExpansion
%EndExpansion

\section{Decay models for $X\left(  3915\right)  \rightarrow D\overline{D}$
\label{SIII}}

In order to get detailed predictions for the open flavor strong decay
$X\left(  3915\right)  \rightarrow D\overline{D}$ we shall rely on a quark
model framework. $X\left(  3915\right)  $ shall be considered as a
$0^{+}\left(  0^{++}\right)  $ $c\overline{c}$ (Cornell or GSPM) state whose
decay takes place through the formation of a light $q\overline{q}$ ($q=u,d$)
pair that combines with $c\overline{c}$ giving rise to $D\overline{D}$. In the
so called $^{3}P_{0}$ decay model \cite{LeY73} $q\overline{q}$ is created in
the hadronic vacuum with $0^{++}$ quantum numbers. In the so called $C^{3}$
(Cornell Coupled-Channel) decay model \cite{Eic80} the $q\overline{q}$
creation is governed by the same potential generating the spectrum. Both
models give a reasonable description of $I(J^{PC})$ charmonium decays
\cite{Eic80,Ono81} below the corresponding first open flavor meson-meson thresholds.

For the $^{3}P_{0}$ as well as for the $C^{3}$ decay model the width for
$X\left(  3915\right)  \rightarrow D\overline{D},$ in the rest frame of
$X\left(  3915\right)  ,$ can be expressed as
\begin{equation}
\Gamma=2\pi\frac{E_{D}E_{\overline{D}}}{M_{X\left(  3915\right)  }}%
k_{D}\left\vert A\right\vert ^{2} \label{widthformula}%
\end{equation}
where $E_{D}$ $\left(  =E_{\overline{D}}\right)  $ is the energy of the $D$
(or $\overline{D}$) meson given by
\begin{equation}
E_{D}=\sqrt{M_{D}^{2}+k_{D}^{2}}=E_{\overline{D}} \label{relenergy}%
\end{equation}
being $k_{D}$ the modulus of the three-momentum of $D$ (or $\overline{D}$) for
which we shall use the relativistic expression%
\begin{equation}
k_{D}=\frac{\sqrt{\left(  M_{X}^{2}-4M_{D}^{2}\right)  }}{2} \label{threemom}%
\end{equation}
and $A$ stands for the decay amplitude.

\bigskip

In the $^{3}P_{0}$ model one has
\begin{equation}
\left\vert A\right\vert _{^{3}P_{0}}^{2}\equiv\gamma^{2}\left\vert
M\right\vert ^{2} \label{3p0amp}%
\end{equation}
where the constant $\gamma$ specifies the strength of the pair creation, and
the expression for $\left\vert M\right\vert ^{2}$ can be derived from
\cite{Ono81} in a straightforward manner (we use the same notation as in this
reference) as
\begin{equation}
\left\vert M\right\vert ^{2}=\frac{1}{32}I\left(  -\right)  ^{2}
\label{MImenos}%
\end{equation}
where
\begin{equation}
I\left(  -\right)  ^{2}=\left\vert
\begin{array}
[c]{c}%
\frac{1}{\hbar^{\frac{9}{2}}}\int_{0}^{\infty}r_{X}^{2}dr_{X}\psi_{X}\left(
r_{X}\right)  \int p^{2}dp\widetilde{u}_{D}\left(  p\right)  \widetilde
{u}_{\overline{D}}\left(  p\right) \\
\left[  -pj_{1}\left(  \frac{pr_{X}}{\hbar}\right)  j_{0}\left(  \frac{m_{c}%
}{\left(  m_{c}+m_{q}\right)  }\frac{kr_{X}}{\hbar}\right)  +\frac{m_{q}%
}{\left(  m_{c}+m_{q}\right)  }kj_{0}\left(  \frac{pr_{X}}{\hbar}\right)
j_{1}\left(  \frac{m_{c}}{\left(  m_{c}+m_{q}\right)  }\frac{kr_{X}}{\hbar
}\right)  \right]
\end{array}
\right\vert ^{2} \label{msquare}%
\end{equation}
$m_{q}$ is the mass of the light quark, $\psi_{X}$ denotes the radial wave
function of $X\left(  3915\right)  $ in configuration space and $\widetilde
{u}_{D}\left(  p\right)  $ stands for radial wave function of $D$ in momentum
space
\begin{equation}
\widetilde{u}_{D}\left(  p\right)  \equiv\sqrt{\frac{2}{\pi}}\int_{0}^{\infty
}r_{D}^{2}dr_{D}\psi_{D}\left(  r_{D}\right)  j_{0}\left(  \frac{pr_{D}}%
{\hbar}\right)  \label{momwfD}%
\end{equation}
calculated from $\psi_{D},$ the radial wave function of $D$ in configuration space.

In order to simplify the calculation we shall approach as usual $\psi
_{D}\left(  r_{D}\right)  $ by a gaussian (the same expression for
$\psi_{\overline{D}}\left(  r_{\overline{D}}\right)  $)%
\begin{equation}
\psi_{D}\left(  r_{D}\right)  =\frac{2}{\pi^{\frac{1}{4}}R_{D}^{\frac{3}{2}}%
}e^{-\frac{r_{D}^{2}}{2R_{D}^{2}}} \label{rwfD}%
\end{equation}
$R_{D}$ can be fixed either variationally or by requiring it to be equal to
the root mean square (rms) radius, obtained from Cornell or the GSPM
descriptions of $D$ (this implies the reset of the value of the coulomb
strength $\zeta$ to get the spectral mass). By using the rms procedure we get
$R_{D}=0.54$ fm. Then the use of the gaussian instead of the Cornell or the
GSPM wave functions hardly makes any difference.

Notice that we have used $k$ in (\ref{msquare}) for the three-momentum of $D$
(or $\overline{D}$) instead of the fixed $k_{D}.$ This will allow us to
analyze the momentum dependence of the amplitude in order to have some idea of
the possible effect of momentum dependent corrections. In this regard we
should keep in mind that\ i) the strength of the pair creation may depend on
momentum and ii) we calculate the amplitude from a non relativistic quark model.

\bigskip

The calculated $^{3}P_{0}$ widths for both descriptions, using $k=k_{D}$, are
plotted in Table~\ref{tabwidthskd}

\begin{center}
\begin{table}[ptb]%
\begin{tabular}
[c]{cc|c}
& $%
\begin{array}
[c]{c}%
X\left(  3915\right) \\
2p\text{ Cornell state}%
\end{array}
$ & $%
\begin{array}
[c]{c}%
X\left(  3915\right) \\
1p_{\left[  T_{1},T_{2}\right]  }\text{ GSPMstate}%
\end{array}
$\\\hline
$%
\begin{array}
[c]{c}%
^{3}P_{0}\\
\text{decay model}%
\end{array}
$ & $1.99\gamma^{2}$ & $0.59\gamma^{2}$\\
&  & \\
$%
\begin{array}
[c]{c}%
C^{3}\\
\text{decay model}%
\end{array}
$ & $33.66$ & $1.95$%
\end{tabular}
\caption{Calculated widths in MeV for the decay $X\left(  3915\right)
\rightarrow D\overline{D}.$ The masses $M_{X\left(  3915\right)  }=3918$ MeV,
$M_{D}=M_{\overline{D}}=1865$ MeV, from \cite{PDG14}, have been used. }%
\label{tabwidthskd}%
\end{table}
\end{center}

\begin{figure}
[ptb]
\begin{center}
\includegraphics[
height=2.2433in,
width=3.659in
]%
{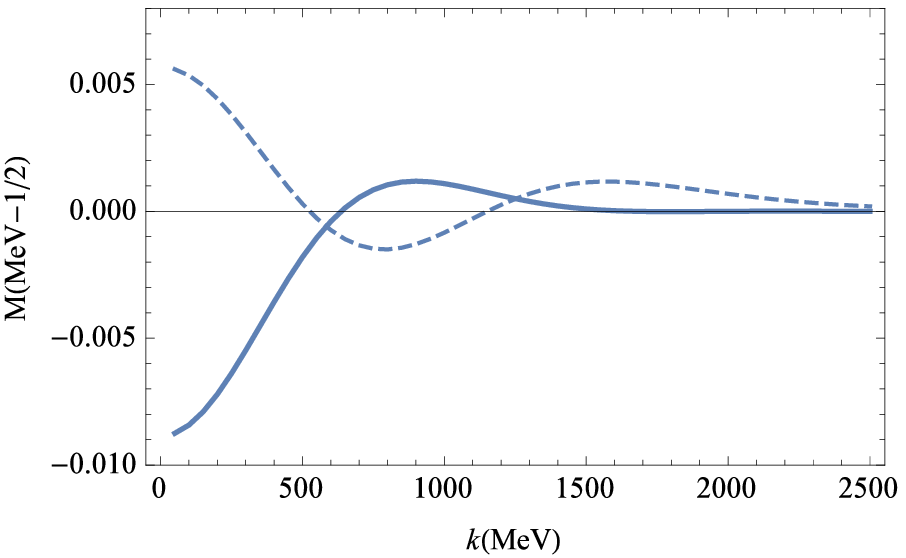}%
\caption{Momentum dependence of the $^{3}P_{0}$ decay ampitude for the
GSPM\ (solid line) and Cornell (dashed line) descriptions of $X\left(
3915\right)  .$}%
\label{momdep3p0}%
\end{center}
\end{figure}

Usual values of $\gamma$, fitted from measured decays, are between $0.4$ and
$7$. Therefore the calculated widths range from a few to dozens of MeV. This
seems to be in contradiction with the observed experimental absence of the
decay. Nonetheless, it is illustrative to examine the momentum dependence of
the amplitude, plotted in Fig. \ref{momdep3p0}.

%

%TCIMACRO{\FRAME{ftbpFU}{3.659in}{2.2433in}{0pt}{\Qcb{Momentum dependence of
%the $^{3}P_{0}$ decay ampitude for the GSPM\ (solid line) and Cornell (dashed
%line) descriptions of $X\left(  3915\right)  .$}}{\Qlb{momdep3p0}%
%}{momdep3P0.eps}{\special{ language "Scientific Word";  type "GRAPHIC";
%width 3.659in;  height 2.2433in;  depth 0pt;  original-width 3.6115in;
%original-height 2.2035in;  cropleft "0";  croptop "1";  cropright "1";
%cropbottom "0";  filename 'momdep3P0.eps';file-properties "XNPEU";}} }%
%BeginExpansion
%EndExpansion

As can be checked, for the GSPM description the amplitude vanishes for a value
of $k=637$ MeV close to $k_{D}=599.6$ MeV. Hence it is plausible that momentum
dependent corrections to the $^{3}P_{0}$ decay model make the amplitude to
vanish. Indeed, it has been shown that the use of harmonic wave functions for
$X\left(  3915\right)  $ as well as for $D$ and $\overline{D}$ gives rise to a
vanishing amplitude \cite{Bar04}.

We may then tentatively conclude that the GSPM description combined with the
$^{3}P_{0}$ decay model might provide an explanation to data.

\bigskip

It should be pointed out that alternative quark model calculations of the
$X\left(  3915\right)  \rightarrow D\overline{D}$ decay from a $^{3}P_{0}$
decay model can be found in the literature. For example, in reference
\cite{Liu10}, with harmonic oscillator wave functions, the estimated width of
the $0^{++}\left(  2p\right)  $ state, close to the total width of $X\left(
3915\right)  $ (a similar value was obtained in \cite{Bar05}), was used as an
argument in favor of a conventional description. In reference \cite{You10},
using a screened potential model description of $X\left(  3915\right)  $
\cite{Li09}, the calculated width was much larger than data disfavoring the
$2p$ state assignment to $X\left(  3915\right)  $. A different result was
found in reference \cite{Jia13}, where the node structure in the Bethe
Salpeter wave function employed gave rise to a small width. Finally, in
reference \cite{Wan14}, by using a gaussian expansion method in the framework
of a chiral quark model to generate the wave functions, a width bigger than
the total width of $X\left(  3915\right)  $ was found.

\bigskip

On the other hand, for the $C^{3}$ decay model one has%
\begin{equation}
\left\vert A\right\vert _{Cor}^{2}\equiv\left\vert G\right\vert ^{2}
\label{C3amp}%
\end{equation}

The expression for $\left\vert G\right\vert ^{2}$ has been derived from
\cite{Eic80} by including the coulomb term of the potential as well as
confinement. By using gaussians wave functions for $D$ and $\overline{D}$ as
above and defining%
\begin{equation}
\beta\equiv\frac{1}{2R_{D}^{2}} \label{betaD}%
\end{equation}
the amplitude reads%
\begin{equation}
\left\vert G\right\vert ^{2}=g^{2}\left\vert \int_{0}^{\infty}dr_{X}\psi
_{X}\left(  r_{X}\right)  e^{-\frac{\beta r_{X}^{2}}{2}}j_{0}\left(
\frac{m_{c}}{\left(  m_{c}+m_{q}\right)  }\frac{kr_{X}}{\hbar}\right)
J\left(  r_{X}\right)  \right\vert ^{2} \label{Gamp}%
\end{equation}
where%
\begin{equation}
g^{2}\equiv\left(  \frac{2}{3\pi^{2}\hbar m_{q}^{2}\beta}\right)  \label{gfac}%
\end{equation}
and%
\begin{equation}
J\left(  r_{X}\right)  \equiv\int_{a}^{b}dre^{-2\beta r^{2}}\left[
2r_{X}\beta\left(  \sigma r+\frac{\zeta}{r}\right)  \left(  e^{-2\beta rr_{X}%
}+e^{2\beta rr_{X}}\right)  +\left(  \sigma+\frac{\zeta}{r^{2}}\right)
\left(  e^{-2\beta rr_{X}}-e^{2\beta rr_{X}}\right)  \right]  \label{jota}%
\end{equation}
For the Cornell description of $X\left(  3915\right)  $ the integration limits
are $a=0$ and $b=\infty$ whereas for the GSPM description one has $a=r_{T_{1}%
}=1.31$ fm and $b=r_{T_{2}}=1.54$ fm (notice that only for this interval the
radial derivative of the Generalized Screened potential from which the
amplitude is calculated \cite{Eic80} does not vanish).

The expression of the amplitude\ for the Cornell description is connected to
the one given by equation (3.37) in \cite{Eic80}, $I_{21}^{0},$ through%
\begin{equation}
I_{21}^{0}=\beta\frac{\left(  G\right)  _{\zeta=0}}{g\sigma} \label{connec}%
\end{equation}

\bigskip

\begin{figure}
[ptb]
\begin{center}
\includegraphics[
height=2.2886in,
width=3.6207in
]%
{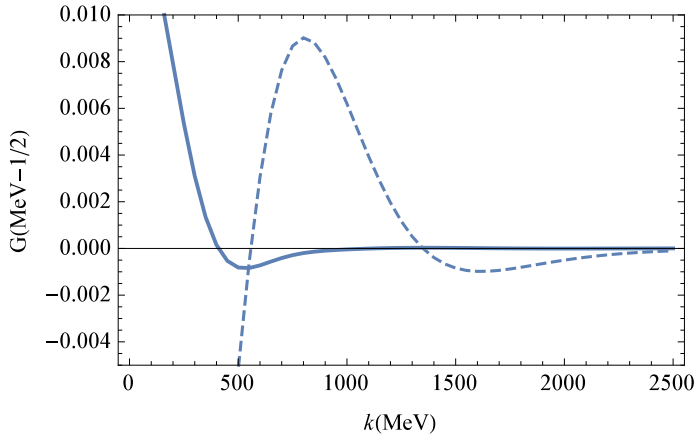}%
\caption{Momentum dependence of the $C^{3}$ decay ampitude for the
GSPM\ (solid line) and Cornell (dashed line) descriptions of $X\left(
3915\right)  .$}%
\label{momdepcornell}%
\end{center}
\end{figure}

The calculated $C^{3}$ decay widths for both descriptions, using $k=k_{D}$,
are plotted in Table~\ref{tabwidthskd}. Again, the values obtained do not fit
data. But if we plot the momentum dependence of the amplitude, Fig.
\ref{momdepcornell}, we realize that for the Cornell description the amplitude
vanishes for a value of $k=558$ MeV close to $k_{D}=599.6$ MeV (this result
differs slightly from the one obtained in reference \cite{Eic04}\ due to the
differences in the expression of the amplitude).%

%TCIMACRO{\FRAME{ftbpFU}{3.6207in}{2.2886in}{0pt}{\Qcb{Momentum dependence of
%the $C^{3}$ decay ampitude for the GSPM\ (solid line) and Cornell (dashed
%line) descriptions of $X\left(  3915\right)  .$}}{\Qlb{momdepcornell}%
%}{momdepcornell.eps}{\special{ language "Scientific Word";  type "GRAPHIC";
%width 3.6207in;  height 2.2886in;  depth 0pt;  original-width 3.8522in;
%original-height 2.4252in;  cropleft "0";  croptop "1";  cropright "1";
%cropbottom "0";  filename 'momdepCornell.eps';file-properties "XNPEU";}} }%
%BeginExpansion
%EndExpansion

Hence it is plausible that momentum dependent corrections to the $C^{3}$ decay
model make the amplitude to vanish. As a matter of fact, the use of
$M_{X}=3910.9$ MeV, as given by the model, and $M_{D}=1869$ MeV as it
corresponds to $D^{+}$ would give a vanishing amplitude for the non
relativistic value of $K_{D}$.

We may then tentatively conclude that the Cornell description combined with
the $C^{3}$ decay model might provide an explanation to data.

\bigskip

Putting together our tentative conclusions we may finally conclude that the
observed suppression of the decay $X\left(  3915\right)  \rightarrow
D\overline{D}$ might be equally well explained from a $C^{3}$ decay
model\ with a Cornell description of $X\left(  3915\right)  $ and from a
$^{3}P_{0}$ decay model with a GSPM description of $X\left(  3915\right)  $.
Therefore, no conclusion about the conventional or unconventional nature of
$X\left(  3915\right)  $ can be extracted from its decay to $D\overline{D}$.

\bigskip

Notice that the experimental suppression of the decay to $D\overline{D}$ may
also be understood, at least qualitatively, from molecular like pictures.
Thus, for instance,\ in reference \cite{Mo09} the dynamically generated
$X\left(  3915\right)  $ (identifying $X(3915)$ with the so called $Y(3940)$
as done by the PDG) was dominantly a $D^{\ast}\overline{D}^{\ast}$ bound state
decaying into pairs of light vectors or light vector-heavy vector mesons,
whereas in reference \cite{Li15} the $X\left(  3915\right)  $ was assumed to
be a $D_{s}\overline{D}_{s}$ bound state so that its decay to $D\overline{D}$
is OZI suppressed.

\section{The $X\left(  3915\right)  \rightarrow\omega J/\psi$ decay
\label{SIV}}

Experimental information on this decay comes from the average of
measured\ production rates in two-photon fusion \cite{PDG14}
\begin{equation}
\Gamma\left(  X(3915)\rightarrow\gamma\gamma\right)  \mathcal{B}\left(
X(3915)\rightarrow J/\psi\omega\right)  =54\pm9\text{ eV} \label{twofbrexp}%
\end{equation}
and from the average of the product of branching fraction measurements for
$X(3915)$ production in $B$ decay (see \cite{Ols15} and references therein)%
\begin{equation}
\mathcal{B}\left(  B^{+}\rightarrow K^{+}X(3915)\right)  \mathcal{B}\left(
X(3915)\rightarrow J/\psi\omega\right)  =3.0_{-0.5-0.3}^{+0.6+0.5}%
\times10^{-5} \label{bfprod}%
\end{equation}

\bigskip

In reference \cite{Ols15} it has been argued that if $X(3915)$ were a
$\chi_{c0}\left(  2p\right)  $ Cornell state then it would be reasonable to
assume that
\begin{equation}
\mathcal{B}\left(  B^{+}\rightarrow K^{+}\chi_{c0}\left(  2p\right)  \right)
\lesssim\mathcal{B}\left(  B^{+}\rightarrow K^{+}\chi_{c0}\left(  1p\right)
\right)  \label{Cornellbound}%
\end{equation}

The argument is based on the fact that the available phase space for
$B^{+}\rightarrow K^{+}\chi_{c0}\left(  2p\right)  $ is significantly smaller
than for $B^{+}\rightarrow K^{+}\chi_{c0}\left(  1p\right)  $ and on the
assumption, based on reference \cite{Bod92}, that $B$-meson decay rate to
$\chi_{c0}\left(  np\right)  $ is proportional to $\left\vert R_{\chi
_{c0}\left(  np\right)  }^{\prime}(0)\right\vert ^{2}$ (we shall discuss this
assumption for conventional Cornell states later on).

\bigskip

As the values of $\left\vert R_{0^{++}\left(  2p\right)  }^{^{\prime}}\left(
0\right)  \right\vert ^{2}$ and $\left\vert R_{0^{++}\left(  1p\right)
}^{^{\prime}}\left(  0\right)  \right\vert ^{2}$ do not differ much (see
(\ref{ratiocornell}) below) one expects the ratio
\begin{equation}
\frac{\mathcal{B}\left(  B^{+}\rightarrow K^{+}\chi_{c0}\left(  2p\right)
\right)  }{\mathcal{B}\left(  B^{+}\rightarrow K^{+}\chi_{c0}\left(
1p\right)  \right)  }\lesssim1 \label{prodratio}%
\end{equation}

Then, using the experimental value $\mathcal{B}\left(  B^{+}\rightarrow
K^{+}\chi_{c0}\left(  1p\right)  \right)  =1.5_{-0.14}^{+0.15}\times10^{-4}$
one would get from (\ref{bfprod})
\begin{equation}
\mathcal{B}\left(  \chi_{c0}\left(  2p\right)  \rightarrow J/\psi
\omega\right)  >0.14 \label{olsbr}%
\end{equation}

\bigskip

On the other hand $\Gamma\left(  X(3915)\rightarrow\gamma\gamma\right)  $ is
known to be proportional to $\left\vert R_{X(3915)}^{\prime}(0)\right\vert
^{2}$ \cite{kwo88}. Therefore, if $X(3915)$ were a $\chi_{c0}\left(
2p\right)  $ Cornell state we would expect the predicted ratio
\begin{equation}
\frac{\left(  \Gamma\left(  \chi_{c0}\left(  2p\right)  \rightarrow
\gamma\gamma\right)  \right)  }{\left(  \Gamma\left(  \chi_{c0}\left(
1P\right)  \rightarrow\gamma\gamma\right)  \right)  }=\frac{\left\vert
R_{0^{++}\left(  2p\right)  }^{^{\prime}}\left(  0\right)  \right\vert ^{2}%
}{\left\vert R_{0^{++}\left(  1p\right)  }^{^{\prime}}\left(  0\right)
\right\vert ^{2}}=1.4 \label{ratiocornell}%
\end{equation}
to be a reasonable approximation to data. Then, using the experimental value
\begin{equation}
\Gamma\left(  \chi_{c0}\left(  1P\right)  \rightarrow\gamma\gamma\right)
=2.3\pm0.4\text{ KeV} \label{c01pgg}%
\end{equation}
one would get%
\begin{equation}
\Gamma\left(  \chi_{c0}\left(  2p\right)  \rightarrow\gamma\gamma\right)
\sim3.3\pm0.6\text{ KeV} \label{c02pgg}%
\end{equation}

However, the combination of (\ref{c02pgg}) with (\ref{twofbrexp}) would give%
\begin{equation}
\mathcal{B}\left(  \chi_{c0}\left(  2p\right)  \rightarrow J/\psi
\omega\right)  \sim0.017\pm0.006 \label{gonbf}%
\end{equation}
which is clearly incompatible with (\ref{olsbr}).

\bigskip

We may then tentatively conclude that the Cornell description of $X(3915)$ is
not consistent with existing data for $\mathcal{B}\left(  X(3915)\rightarrow
J/\psi\omega\right)  .$

\bigskip

Let us now consider the GSPM description, say $X(3915)$ is a $1p_{\left[
T_{1},T_{2}\right]  }$ GSPM state. By using again the assumption
$\mathcal{B}\left(  B^{+}\rightarrow K^{+}\chi_{c0}\left(  2p\right)  \right)
\lesssim\mathcal{B}\left(  B^{+}\rightarrow K^{+}\chi_{c0}\left(  1p\right)
\right)  $ an upper bound for $\mathcal{B}\left(  B^{+}\rightarrow
K^{+}X_{1p_{\left[  T_{1},T_{2}\right]  }}\right)  $ can be found as follows.
From \cite{Bod92} the decay rate of a $B^{+}$-meson to $0^{++}$ charmonium is
given by the decay rate of the $\overline{b}$ antiquark with the light quark
as a noninteracting spectator. To leading order in the QCD coupling the
production rate, involving a color octet mechanism (a $c\overline{c}$ pair
produced in a color octet $S$-wave), can be written as%
\begin{equation}
\Gamma_{\left(  \overline{b}\rightarrow0^{++},\overline{s}\right)  }%
=H_{8}^{\prime}\left(  m_{b}\right)  \Gamma_{8\left(  \overline{b}\rightarrow
c\overline{c}\left(  ^{3}S_{1}\right)  ,\overline{s}\right)  } \label{prodb}%
\end{equation}
where the subindex $8$ stands for color octet mechanism, $m_{b}$ is the mass
of the $b$ quark and $H_{8}^{\prime}\left(  m_{b}\right)  $ is a
nonperturbative parameter proportional to the probability for a $c\overline
{c}$ pair produced in a color octet $S$-wave fragmenting into a color singlet
$0^{++}$ bound state. This parameter can be expressed as%
\begin{equation}
H_{8}^{\prime}\left(  m_{b}\right)  =a+eH_{1} \label{hprim}%
\end{equation}
where$\ a$ is an unknown constant to be determined phenomenologically, $H_{1}$
is given by
\begin{equation}
H_{1}\approx\frac{1}{m_{c}^{4}}\left(  \frac{9}{2\pi}\right)  \left\vert
R_{0^{++}}^{\prime}(0)\right\vert ^{2} \label{h1}%
\end{equation}
with $m_{c}$ the mass of the $c$ quark and
\begin{equation}
e\equiv-\left(  \frac{16}{27\beta}\right)  \ln\left(  \alpha_{s}\left(
m_{b}\right)  \right)  \label{eexpr}%
\end{equation}
with $\beta=\frac{33-2n_{f}}{6}$ being $n_{f}$ the number of active quarks.
Using $n_{f}=4$ and $\alpha_{s}\left(  m_{b}\right)  \approx0.2$ \cite{Bod92}
we get $e\approx0.2$.

\bigskip

As the phase space is the same for the GSPM\ and the Cornell descriptions we
get%
\begin{equation}
\frac{\mathcal{B}\left(  B^{+}\rightarrow K^{+}X_{1p_{\left[  T_{1}%
,T_{2}\right]  }}\right)  }{\mathcal{B}\left(  B^{+}\rightarrow K^{+}\chi
_{c0}\left(  2p\right)  \right)  }=\frac{a+e\left(  H_{1}\right)
_{1p_{\left[  T_{1},T_{2}\right]  }}}{a+e\left(  H_{1}\right)  _{\chi
_{c0}\left(  2p\right)  }} \label{ratioGSPM}%
\end{equation}

By substituting the calculated values
\begin{equation}
\left\vert R_{0^{++}(1p_{\left[  T_{1},T_{2}\right]  })}^{^{\prime}}\left(
0\right)  \right\vert ^{2}=3.57\text{ fm}^{-\frac{5}{2}} \label{derGSPM}%
\end{equation}%
\begin{equation}
\left\vert R_{0^{++}\left(  2p\right)  }^{^{\prime}}\left(  0\right)
\right\vert ^{2}=272.25\text{ fm}^{-\frac{5}{2}} \label{derCornell}%
\end{equation}
we have
\begin{equation}
\frac{\mathcal{B}\left(  B^{+}\rightarrow K^{+}X_{1p_{\left[  T_{1}%
,T_{2}\right]  }}\right)  }{\mathcal{B}\left(  B^{+}\rightarrow K^{+}\chi
_{c0}\left(  2p\right)  \right)  }=\frac{a+0.1\text{ MeV}}{a+7.1\text{ MeV}}
\label{ratioa}%
\end{equation}

\bigskip

Then using $\mathcal{B}\left(  B^{+}\rightarrow K^{+}\chi_{c0}\left(
2p\right)  \right)  \lesssim\mathcal{B}\left(  B^{+}\rightarrow K^{+}\chi
_{c0}\left(  1p\right)  \right)  $ we obtain from (\ref{ratioa}) the bound%
\begin{equation}
\mathcal{B}\left(  B^{+}\rightarrow K^{+}X_{1p_{\left[  T_{1},T_{2}\right]  }%
}\right)  \lesssim\left(  \frac{a+0.1\text{ MeV}}{a+7.1\text{ MeV}}\right)
\mathcal{B}\left(  B^{+}\rightarrow K^{+}\chi_{c0}\left(  1p\right)  \right)
\label{firstbound}%
\end{equation}
in terms of the unknown constant $a$.

\bigskip

Let us consider now $\Gamma\left(  1p_{\left[  T_{1},T_{2}\right]
}\rightarrow\gamma\gamma\right)  .$ This width can be calculated from the
predicted GSPM ratio ((notice that there is no difference between the Cornell
$\chi_{c0}\left(  1P\right)  $ state and the $0^{++}\left(  1p_{\left[
T_{0},T_{1}\right]  }\right)  $ GSPM state)%
\begin{equation}
\frac{\Gamma\left(  1p_{\left[  T_{1},T_{2}\right]  }\rightarrow\gamma
\gamma\right)  }{\Gamma\left(  \chi_{c0}\left(  1P\right)  \rightarrow
\gamma\gamma\right)  }=\frac{\left\vert R_{0^{++}\left(  1p_{\left[
T_{1},T_{2}\right]  }\right)  }^{^{\prime}}\left(  0\right)  \right\vert ^{2}%
}{\left\vert R_{0^{++}\left(  1p_{\left[  T_{0},T_{1}\right]  }\right)
}^{^{\prime}}\left(  0\right)  \right\vert ^{2}}=0.02 \label{gspmratio}%
\end{equation}
Using the experimental value $\Gamma\left(  \chi_{c0}\left(  1P\right)
\rightarrow\gamma\gamma\right)  =2.3\pm0.4$ KeV one gets%
\begin{equation}
\Gamma\left(  0^{++}\left(  1p_{\left[  T_{1},T_{2}\right]  }\right)
\rightarrow\gamma\gamma\right)  \simeq0.02\left(  \Gamma\left(  \chi
_{c0}\left(  1p\right)  \rightarrow\gamma\gamma\right)  \right)  _{Exp}%
=46\pm8\text{ }eV \label{gspmgg2}%
\end{equation}

Therefore, if $X(3915)$ is a $1p_{\left[  T_{1},T_{2}\right]  }$ GSPM state,
the combination of (\ref{gspmgg2}) with (\ref{twofbrexp}) gives
\begin{equation}
\mathcal{B}\left(  1p_{\left[  T_{1},T_{2}\right]  }\rightarrow J/\psi
\omega\right)  >0.83 \label{gspmbf}%
\end{equation}
This implies from (\ref{bfprod}) that%
\begin{equation}
\mathcal{B}\left(  B^{+}\rightarrow K^{+}X_{1p_{\left[  T_{1},T_{2}\right]  }%
}\right)  <3.6_{-0.6-0.4}^{+0.7+0.6}\times10^{-5} \label{gammabound}%
\end{equation}
Hence making this bound equal to the one previously obtained (\ref{firstbound}%
), we get a phenomenological value for $a$ compatible with data. For the
central experimental value $\mathcal{B}\left(  B^{+}\rightarrow K^{+}%
X_{1p_{\left[  T_{1},T_{2}\right]  }}\right)  <3.6\times10^{-5}$ we have%

\begin{equation}
a\sim2.1\text{ MeV} \label{ainterval}%
\end{equation}

Hence a full consistent description of data is feasible. Furthermore, this
value of $a$ gives a ratio%
\begin{equation}
\frac{a+e\left(  H_{1}\right)  _{\chi_{c0}\left(  2p\right)  }}{a+e\left(
H_{1}\right)  _{\chi_{c0}\left(  1p\right)  }}\sim\frac{2.1+7.1}{2.1+5.0}=1.3
\label{rationew}%
\end{equation}

very close to the ratio
\begin{equation}
\frac{\left\vert R_{0^{++}\left(  2p\right)  }^{^{\prime}}\left(  0\right)
\right\vert ^{2}}{\left\vert R_{0^{++}\left(  1p\right)  }^{^{\prime}}\left(
0\right)  \right\vert ^{2}}=1.4 \label{ratioagain}%
\end{equation}

providing consistency to the argument used in \cite{Ols15}.

\bigskip

We may then tentatively conclude that the GSPM description of $X(3915)$ might
be consistent with existing data for $\mathcal{B}\left(  X(3915)\rightarrow
J/\psi\omega\right)  .$

\bigskip

Certainly one could argue that the calculated values of the square of the
derivatives of the wave functions at the origin, on which our discussion is
based, could vary when corrections to the Cornell and\ GSPM descriptions were
considered. However, as we only deal with ratios involving such derivatives we
do not expect significant changes from these corrections.

\bigskip

As the plausible account of data by the GSPM depends on the value of the
unknown parameter $a$ some direct estimation of the $X(3915)\rightarrow
J/\psi\omega$ decay width would be of great interest to confirm or refute our
results. Unfortunately the QCDME (QCD Multipole Expansion) formalism developed
to calculate hadronic decays \cite{Got78,Yan80} is not very reliable when
states above threshold are involved (see for example \cite{Bra11}).
Nonetheless, if we assumed that corrections could be effectively incorporated
by means of multiplicative factors then we might try to apply the QCDME to
compare the decay widths obtained with the GSPM and Cornell descriptions. Even
so, the calculation of these decay widths would be out of the scope of this
article since it involves contributions from intermediate color octet states
that should be consistently obtained with the model under consideration.

The only simple thing we can do is to use a scaling law as a very rough
approach for the ratio of the decay widths being aware that the value obtained
could differ even orders of magnitude from the real one (see \cite{kua09} and
references therein).

In the QCDME the $X(3915)\rightarrow J/\psi\omega$ decay corresponds to a
three gluon E1-E1-E1 transition. As each E1 introduces a color-electric dipole
moment that goes linearly with the $c\overline{c}$ distance $\overrightarrow
{r}$ the scaling law reads
\begin{equation}
\frac{\Gamma\left(  1p_{\left[  T_{1},T_{2}\right]  }\rightarrow J/\psi
\omega\right)  }{\Gamma\left(  \chi_{c0}\left(  2p\right)  \rightarrow
J/\psi\omega\right)  }\simeq\left(  \frac{\int drr^{2}R_{J/\psi}%
r^{3}R_{1p_{\left[  T_{1},T_{2}\right]  }}}{\int drr^{2}R_{J/\psi}r^{3}%
R_{\chi_{c0}\left(  2p\right)  }}\right)  ^{2} \label{scaling}%
\end{equation}
where $R$ stands for the radial wave function.

By substituting the calculated integrals from the GSPM\ and the Cornell
descriptions we get%
\begin{equation}
\frac{\Gamma\left(  1p_{\left[  T_{1},T_{2}\right]  }\rightarrow J/\psi
\omega\right)  }{\Gamma\left(  \chi_{c0}\left(  2p\right)  \rightarrow
J/\psi\omega\right)  }\simeq4 \label{result}%
\end{equation}

This value is about $13$ times smaller than the one obtained from
(\ref{gspmbf}) and (\ref{gonbf}), may be indicating the inadequacy of the
scaling law and signaling the need for a more precise direct estimation of the
$X(3915)\rightarrow J/\psi\omega$ decay before extracting any definite
conclusion about the validity of the GSPM to describe the $X(3915).$

In this regard, it may be also illustrative to compare the\ result
$\Gamma\left(  X(3915)_{GSPM}\rightarrow\gamma\gamma\right)  \simeq46\pm8$ eV,
from which the bound $\mathcal{B}\left(  X(3915)_{GSPM}\rightarrow
J/\psi\omega\right)  >0.83$ is obtained, with those obtained from molecular
like model descriptions of $X(3915)$ (identifying $X(3915)$ with the so called
$Y(3940)$ as done by the PDG)$.$ So, in reference \cite{Bra09}, where the
$X(3915)$ is considered as a $D^{\ast}\overline{D}^{\ast}$ hadronic molecule
and a phenomenological lagrangian approach is followed, the values
$\Gamma\left(  X(3915)\rightarrow\gamma\gamma\right)  \simeq330$ eV and
$\Gamma\left(  X(3915)\rightarrow J/\psi\omega\right)  \simeq5.47$ MeV have
been reported; these values are close to satisfy the experimental requirement
(\ref{twofbrexp}). On the other hand, in reference \cite{Mo09} the calculated
values $\Gamma\left(  X(3915)\rightarrow\gamma\gamma\right)  \simeq31$ eV and
$\Gamma\left(  X(3915)\rightarrow J/\psi\omega\right)  \simeq1.52$ MeV are far
from satisfying (\ref{twofbrexp}). Therefore, the GSPM result for
$\Gamma\left(  X(3915)\rightarrow\gamma\gamma\right)  $ is closer to those
obtained from molecular approaches than to the one resulting from the
conventional Cornell description; with respect to the $X(3915)$ decay to
$J/\psi\omega$ its GSPM value has to be, as shown above, significantly more
dominant than in such approaches in order to reproduce current data.

\section{Summary \label{SV}}

A comparative study of the strong decays $X\left(  3915\right)  \rightarrow
D\overline{D}$ and $X(3915)\rightarrow J/\psi\omega$ has been carried out from
two quark model descriptions of $X\left(  3915\right)  .$ The first
description comes out from a Cornell potential that provides a reasonable fit
to heavy quarkonia states lying below open flavor two meson thresholds; the
second one is based on a Generalized Screened Potential Model (GSPM) that
allows for a consistent heavy quarkonia description of $J^{++}$ states below
and above thresholds and $1^{--}$ states quite below their corresponding
threshold (in this last case there is no difference between the GSPM\ and
Cornell potentials).

The $X\left(  3915\right)  \rightarrow D\overline{D}$ process has been studied
from two decay models, the $^{3}P_{0}$ and the $C^{3}$ (Cornell
Coupled-Channel), usually employed within the quark model framework. We have
shown that the three-momentum of the final mesons ($D$ and $\overline{D}$) is
close to the value that makes $i)$ the $^{3}P_{0}$ decay amplitude to vanish
for a GSPM\ description of $X\left(  3915\right)  $ and $ii)$ the $C^{3}$
decay amplitude to vanish for a Cornell\ description of $X\left(  3915\right)
.$ These results make plausible an explanation of the observed experimental
absence of this decay through small momentum dependent corrections to the
amplitudes. As a consequence, no discrimination between the two descriptions
employed can be done from this decay.

A different situation may occur for $X(3915)\rightarrow J/\psi\omega.$ We have
shown that an explanation of existing data involving the branching fraction
$\mathcal{B}\left(  X(3915)\rightarrow J/\psi\omega\right)  $ seems to be
impossible to attain from the Cornell description. On the contrary, the GSPM
description might accommodate all the experimental information predicting a
quite big branching ratio for this OZI non allowed decay. The experimental
confirmation of this prediction would clearly point out to a non conventional
nature of $X(3915)$ putting in question the $\chi_{c0}\left(  2p\right)  $ PDG assignment.

\bigskip

This work has been supported by Ministerio de Econom\'{\i}a y Competitividad
of Spain (MINECO) grant FPA2013-47443-C2-1-P, by SEV-2014-0398 and by
PrometeoII/2014/066 from Generalitat Valenciana.

\end{document}